\documentclass[8pt]{article}
\usepackage{arxiv}
\usepackage[utf8]{inputenc} 
\usepackage[T1]{fontenc}    
\usepackage{hyperref}       
\usepackage{url}            
\usepackage{booktabs}       
\usepackage{amsfonts}       
\usepackage{nicefrac}       
\usepackage{microtype}      
\usepackage{graphicx}		
\usepackage{amsmath}
\usepackage{amssymb}
\usepackage{widetext}
\usepackage[utf8]{inputenc}
\title{An underground Sagnac gyroscope with sub-prad/s rotation rate sensitivity: toward General Relativity tests on Earth}

\author{
 Angela D.V.~Di~Virgilio, Filippo Bosi, Umberto Giacomelli, Andrea Simonelli and Giuseppe Terreni\\
  INFN Sez. di Pisa,\\
   Polo Fibonacci, Largo B Pontecorvo 3, I-56127 Pisa, Italy \\
\texttt{angela.divirgilio@pi.infn.it} \\
\And
Andrea Basti, Nicol\`o Beverini, Giorgio Carelli, Donatella Ciampini, Francesco Fuso,\\ Enrico Maccioni ans Paolo Marsili\\
  Universit\`a di Pisa,\\ Dipartimento di Fisica "E. Fermi", Largo B Pontecorvo 3, I-56127 Pisa, Italy
 \And
Antonello Ortolan\\
       INFN-National Lab. of Legnaro, viale dell'Universit\`a 2, I-35020, Legnaro (PD), Italy \\
       \And
       Alberto Porzio\\
       CNR-SPIN and INFN, Napoli, Complesso Univ.\ Monte Sant'Angelo, via Cintia, Napoli, Italy
}

\begin{document}
\maketitle

\begin{abstract}
Measuring in a single location on Earth its angular rotation rate with respect to the celestial frame, with a sensitivity enabling access to the tiny Lense-Thirring effect  is an extremely challenging task.
GINGERINO is a large frame ring laser gyroscope, operating free running and unattended inside the underground laboratory of the Gran Sasso, Italy. The main geodetic signals, i.e., Annual and Chandler wobbles, daily polar motion and Length of the Day, are recovered from GINGERINO data using standard linear regression methods, demonstrating a sensitivity better than 1 prad/s, therefore close to the requirements for an Earth-based Lense-Thirring test.
\end{abstract}
\maketitle

\begin{figure}
    \centering
    \includegraphics[scale=0.8]{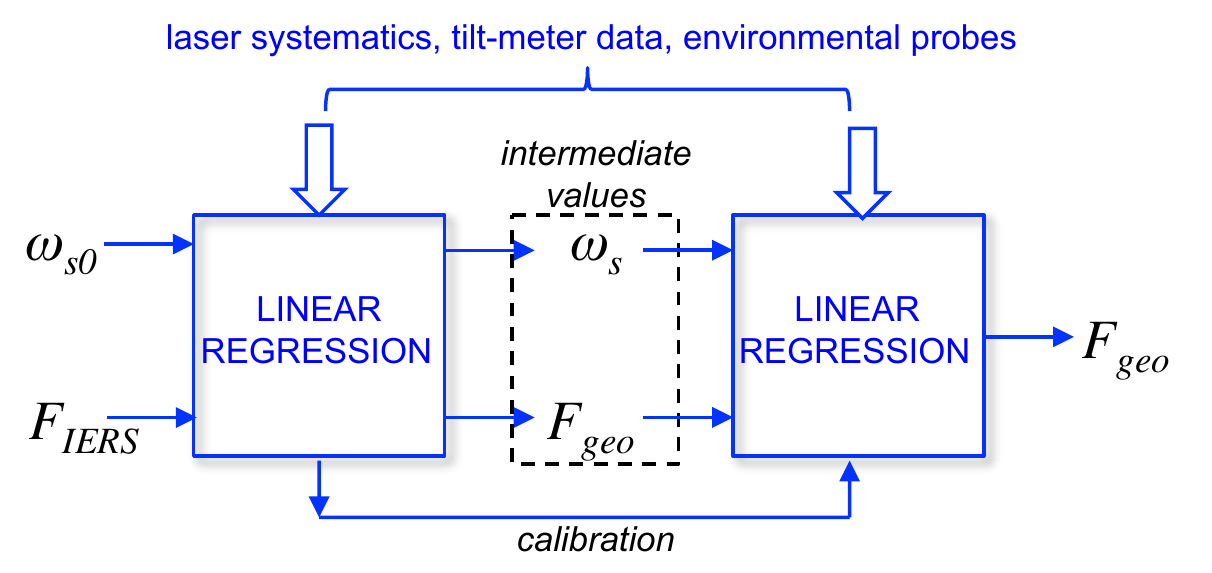}
    \caption{Sketch of the linear regression procedure. In the first step the initial evaluation of the Sagnac angular frequency $\omega_{s0}$  and the $F_{IERS}$ data represent the input of the linear regression, along with a vector containing all terms related to laser systematics, tilt-meter signals, and environmental probes. An intermediate value of $\omega_{s}$ and an intermediate estimate of $F_{geo}$ are obtained at the output, along with a factor accounting for the cross-calibration of the Sagnac frequency with $F_{IERS}$. In the subsequent step the refined estimate of $F_{geo}$ is obtained, accounting for cross-calibration. The procedure is iterated up to convergence. }
    \label{Scheme}
\end{figure}

\begin{figure}
    \centering
    \includegraphics[scale=0.25]{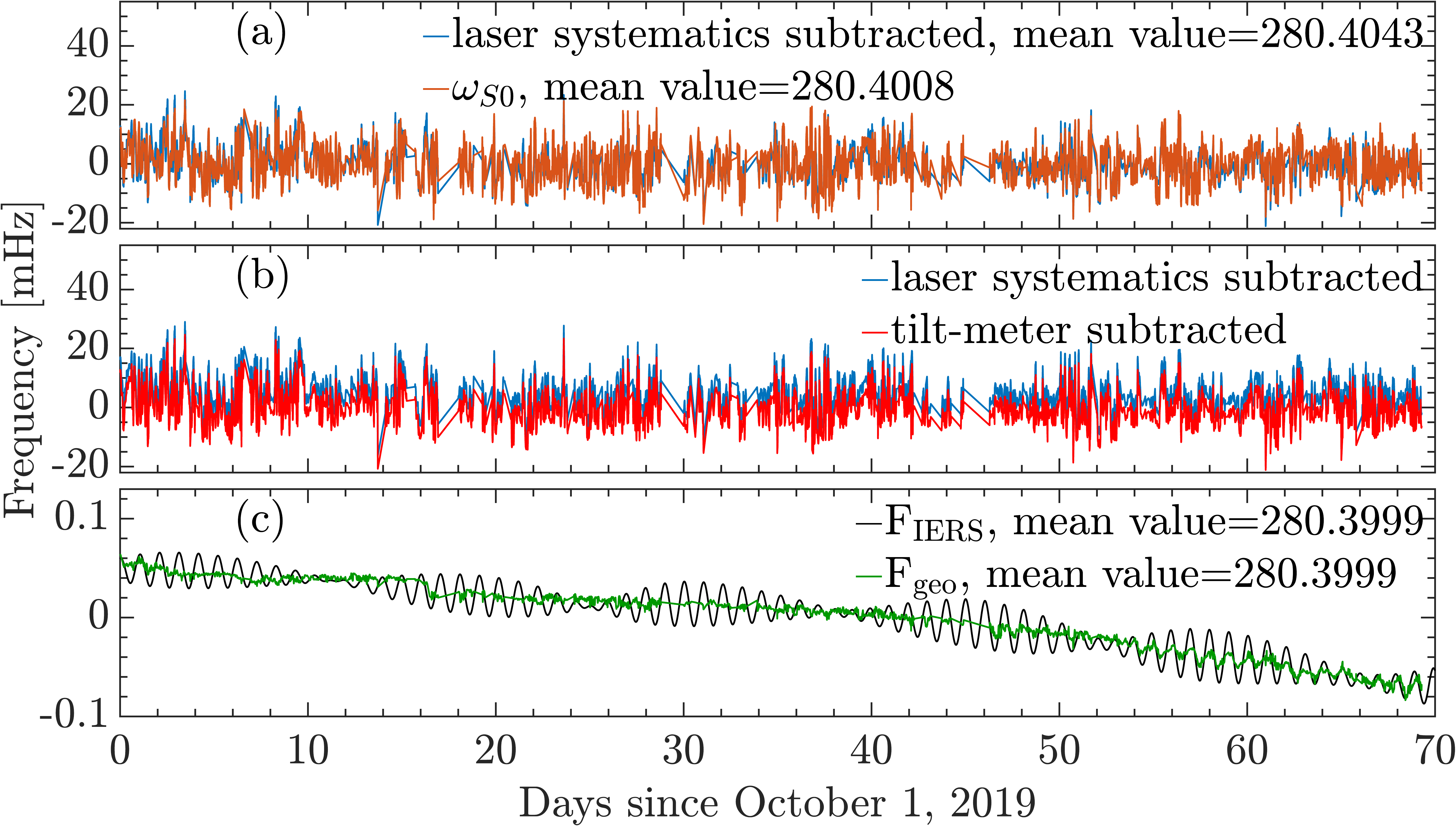}
    \caption{
    Progress of the linear regression procedure (single-fit method): 
    $\omega_{s0}$ and $F_{eff}$ obtained by subtracting effects related to laser systematics (a); $F_{eff}$ after further subtraction of the effects related to tilt-meter signals (b);
    the evaluated $F_{geo}$ compared with $F_{IERS}$ (c). In order to better show the sensitivity of the measurement, plotted data are subtracted for the mean values, as reported in the legends in Hz. Note that only data selected for the analysis are shown in the graph, being missing data replaced by straight lines.}
    \label{fig:sub70}
\end{figure}

\begin{figure}
    \centering
    \includegraphics[scale=0.25]{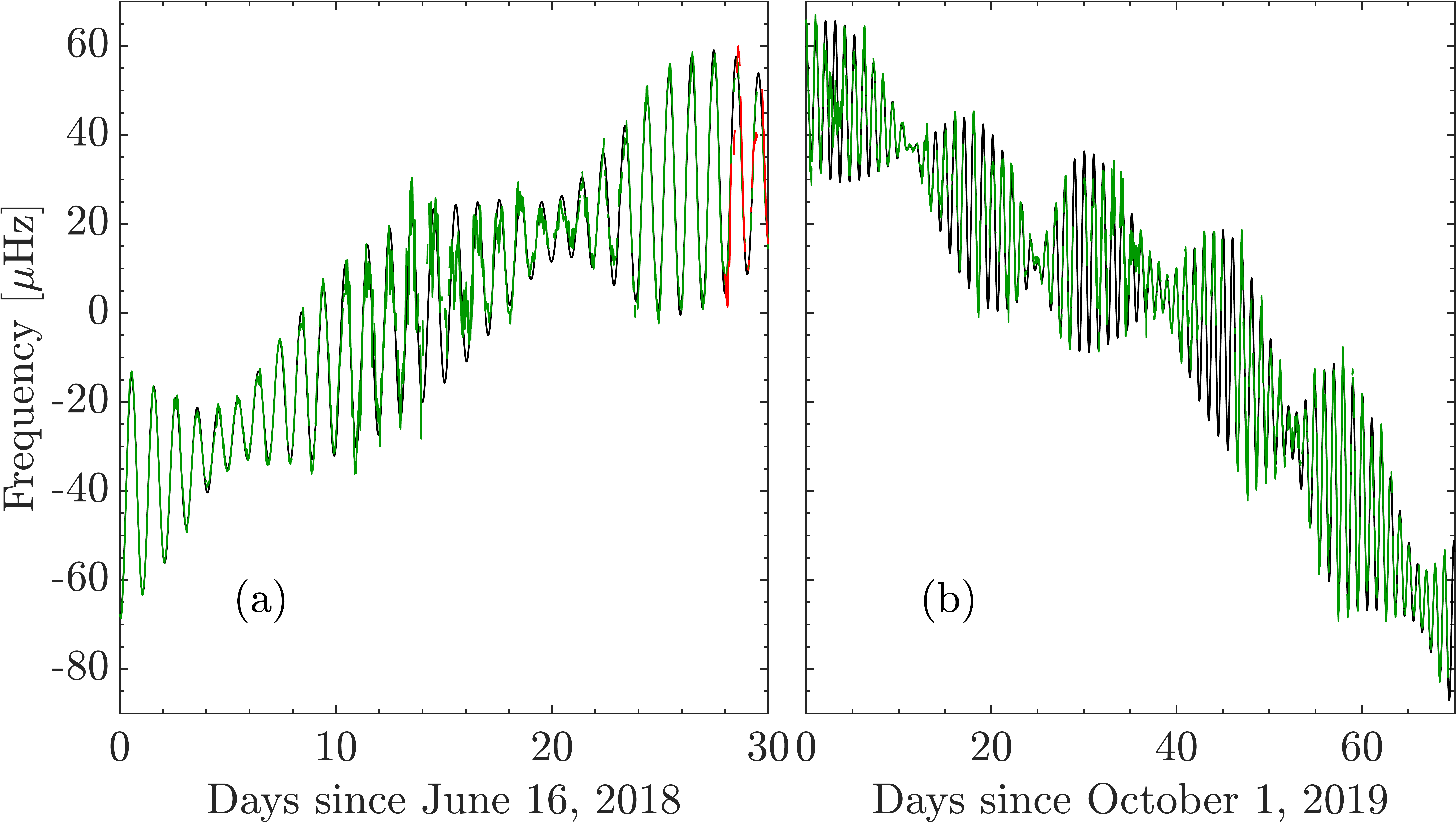}
    \caption{
    The evaluated $F_{geo}$ compared with $F_{IERS}$ (blue and green lines, respectively) for the 2018 (a) and the 2019 (b) data sets. Evaluation is carried out with the moving window method. Data marked in red in panel (a) demonstrates the predictive abilities of the approach, as discussed in the text. In both panels mean values have been subtracted.}
    \label{fig:comp70}
\end{figure}

\begin{figure}
\includegraphics[scale=0.3]{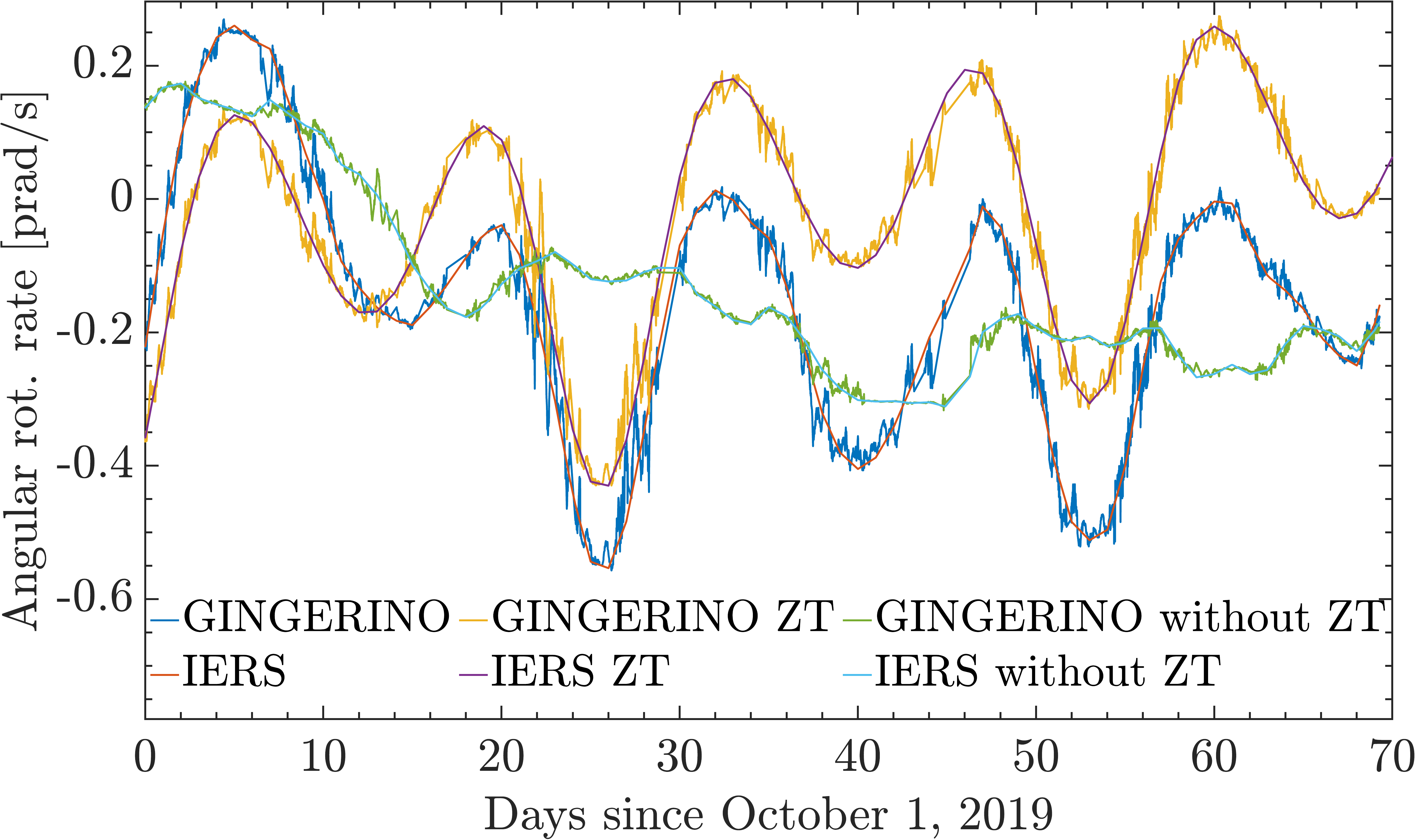}
   \caption{ Effects of the low frequency variations $\Delta \omega_3$ on the Earth angular rotation rate evaluated from GINGERINO data and compared with IERS data.  The comparison is carried out  both including and neglecting contributions from zonal tides, as discussed in the text. The 2019 data set is considered.
  }
    \label{fig:lod1}
\end{figure}

\begin{figure}
    \centering
 \includegraphics[scale=0.3]{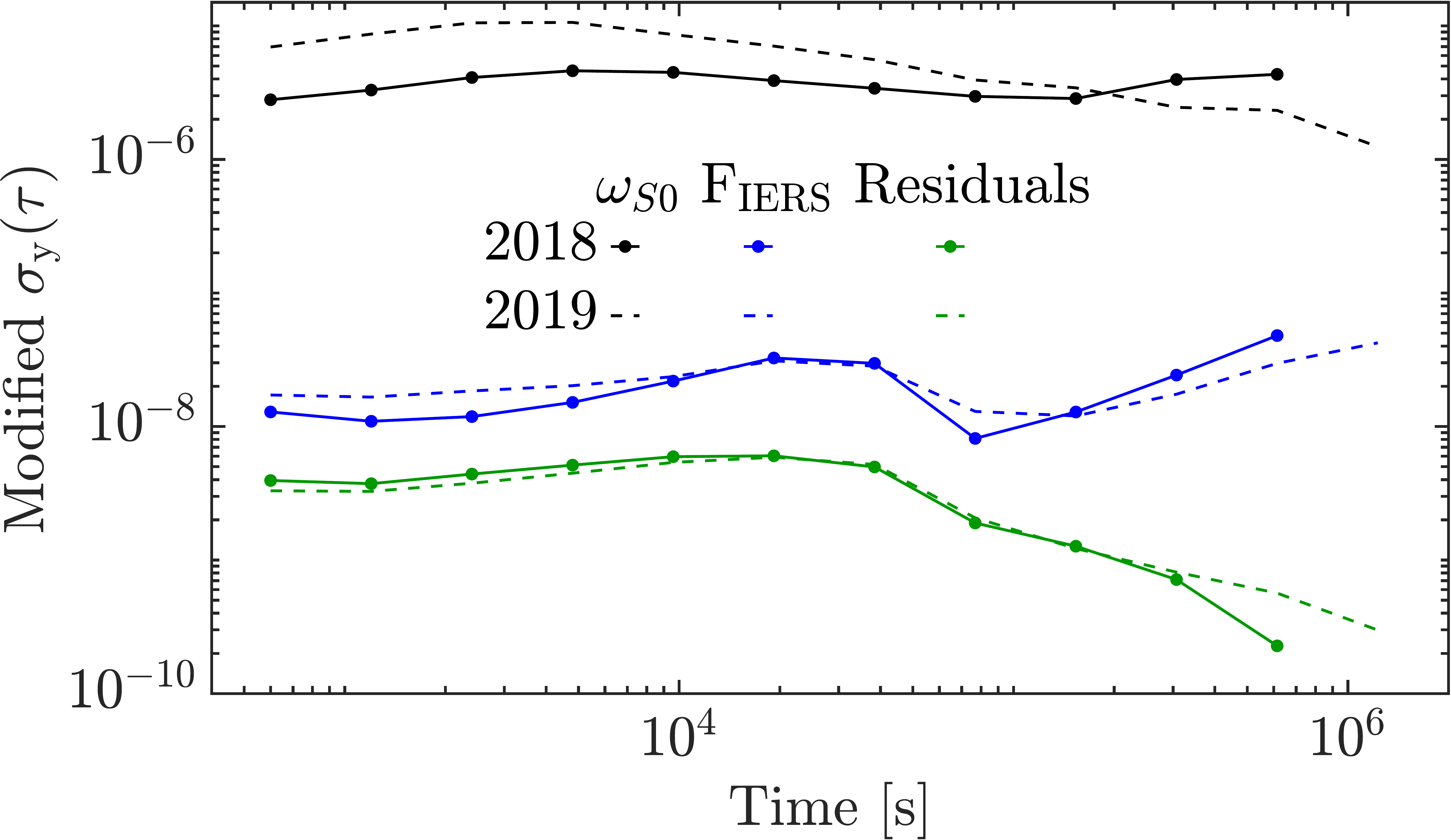}
    \caption{ Modified Allan Deviation $\sigma_y(\tau)$ calculated for $\omega_{s0}$, for $F_{geo}$, and for the residuals defined as in the text. Both 2019 (dotted lines) and 2018 (solid lines) data sets are shown. }
    \label{fig:MADl}
\end{figure}

Sensing rotation rate is essential for both applications and fundamental science. Ring Laser Gyroscopes (RLGs), based on the Sagnac effect, have been established as top sensitivity instruments for measuring rotation rates relative to an inertial frame with an excellent accuracy.\cite{Schreiber2013} Further to geodetic information relating to its instantaneous rotation, knowing the absolute value of the Earth rotation rate and investigating its variations with an Earth-based instrument  are of paramount interest to detect relativistic effects, e.g., the Lense-Thirring
one.\cite{ciuf:nature,GPB,PRD:2011,Tartaglia2017a,Angela2017,Lucchesi_2015} Improving accuracy and reliability in data analysis is a crucial point for enabling the use of RLGs in General Relativity (GR) Earth-based measurements and in the investigation of new physics theories.\cite{CAPOZZIELLO2011167,Jay2019,Aldrovandi:2013wha,CLIFTON20121}
Among geodetic effects, the  polar motion, mainly composed of daily variations, and the Annual and Chandler wobbles have been already observed by monolithic RLG.\cite{Schreiber2011a} Polar motion has been analytically modelled in terms of its local effects as a function of time, latitude, and longitude on a RLG.\cite{Tercjak2017} The wobbles are routinely and constantly measured by IERS (International Earth Rotation and reference system Service) on a daily basis.\cite{IERS}
Moreover, variations of the Length of the Day (LOD) affect the Earth angular rotation rate $\Omega_\oplus$ via a term defined as $\Delta\omega_3$. Therefore, the Sagnac frequency $f_s$ measured by an RLG is affected by both variations of the absolute value of angular rotation rate and of its projection, which cannot be disentangled each other by a single RLG.\cite{Angela2017} RLGs are however considered the only instruments able to provide almost real-time sub-daily measurements of the relevant quantities.\cite{Schreiber2013,G2019}
\\
In this Letter we demonstrate that the sensitivity of an heterolithic RLG, GINGERINO, can be pushed to the envelope of the GR sensitivity region by applying statistical methods to the analysis of its data. GINGERINO \cite{Belfi:17} is a $3.6~$m-side RLG in continuous, unattended operation inside the underground Gran Sasso laboratory (LNGS, Italy) and able to provide data with a duty cycle  $\gtrsim 80\%$.\cite{Note1,Belfi:17,Belfi:18}
Purpose of the statistical methods is to look
for evidence of geodetic signals, $F_{geo}$, in the acquired data. 
A linear regression procedure evaluates the parameters weighting contribution of different signals accounting for laser dynamics, information from a co-located tilt-meter and environmental probes (local temperature and tides), in order to find the best estimate of the expected geodetic signals $F_{IERS}$.
\\
The signal of interest in the analysis is the Sagnac frequency $f_s$, which is proportional to the total angular velocity $\vec{\Omega}_T$  according to \cite{Tartaglia2017a, Angela2017}
\begin{eqnarray}
     f_s = S\times\Omega_T \cos{\beta} \label{general} \\
     S = 4\frac{A}{L\lambda},\nonumber
 \end{eqnarray}
 where $S$ is the scale factor depending on the area $A$ enclosed in the cavity, its perimeter $L$, and the laser wavelength $\lambda$, while $\beta$ is the angle between the area vector and $\vec{\Omega}_T$.
 $\vec{\Omega}_T$ is the total angular velocity with respect to an inertial frame experienced by the gyroscope optical cavity, resulting from the sum of several terms: the dominant one, the Earth angular rotation rate $\vec{\Omega}_{\oplus}$, and the local and instrumental rotations, in principle unknown, defined as $\vec{\omega}_L$.
The Sagnac frequency $f_s$ is also sensitive to fluctuations of the angle $\beta$ due to polar motion and to local and instrumental tilting.
 Accordingly we can consider the Earth angular rotation $\Omega_\oplus =\overline{\Omega}_\oplus +\Delta\omega_3 + \Omega_{GR}$, 
where $\overline{\Omega}_\oplus$ indicates the nominal value,\cite{IERS} $\Delta\omega_3$ accounts for low frequency variations of the Earth rotation rate connected to changes of LOD and to zonal tides, and $\Omega_{GR}$ the GR effects, due to the fact that GINGERINO  is connected to the Earth crust and operates in a rotating non inertial frame. At the latitude of the Gran Sasso underground laboratory, for a RLG lying in a horizontal plane the effect of $\Omega_{GR}$ is expected to be $2.29\times 10^{-14}$ rad/s, corresponding, for the case of GINGERINO, to a shift in $f_s$ of $0.130$ $\mu$Hz.\cite{Tartaglia2017a}
 Instrumental and local effects have to be integrated in the analysis. Assuming the scale factor $S$ constant, reasonable in a first approximation owing to the large temperature stability of the underground laboratory (typical amplitude fluctuations are of the order of $0.1~{^\circ}$C in one month) and the small thermal expansion of the granite-made gyroscope frame ($6.5\times 10^{-6}~^\circ$C$^{-1}$), instrumental and local changes in rotation velocity $\omega_L$ and variations of the absolute orientation $\delta \theta_L$ can be included in Eq.~\ref{general}. First order expansion leads to an effective Sagnac frequency $F_{eff}$:
  \begin{eqnarray}\label{effective}
    F_{eff} &=& S [\sin\theta (\Delta\omega_3 +\omega_L+\overline{\Omega}_\oplus +\Omega_{GR})+\\
&& +\cos\theta (\delta \theta_L+PM) (\Delta\omega_3 +\omega_L+\overline{\Omega}_\oplus +\Omega_{GR})],\nonumber
 \end{eqnarray}
 where $PM$ represents combined effects of polar motion and of the Annual and Chandler wobbles. The first term in Eq.~\ref{effective} depends on actual changes of angular velocity, while the second one is due to changes in the projection. $F_{eff}$ can be decomposed as a sum of $F_{IERS}$ and of the local $F_L$:
  \begin{eqnarray}
  F_{IERS} &=& S~[(\Delta\omega_3 +\overline{\Omega}_\oplus +\Omega_{GR})\sin\theta+\\&& PM (\Delta\omega_3 +\overline{\Omega}_\oplus +\Omega_{GR}) \cos\theta ]\nonumber \label{basicTheory}\\
  \label{IERS}
  F_L &=& S~[ \omega_L\sin\theta +PM  \omega_L\cos\theta \\
  && +\delta \theta_L (\Delta\omega_3+\omega_L+\overline{\Omega}_\oplus+\Omega_{GR}) \cos\theta].\nonumber
 \label{proj}
  \end{eqnarray}
Using the available data, $F_{IERS}$ has been evaluated for the gyroscope location.\cite{PrivMonika}
The main purpose of the linear regression is to compare the Sagnac frequency of the gyroscope with the expected $F_{IERS}$, identifying the local signals $F_L$ by means of the available environmental probes. Evaluation of $\delta\theta_L$ is based on the two-channel tilt-meter located on top of the RLG monument. Using both channels ($\delta\theta_{L1}$ and $\delta\theta_{L2}$) enables accounting for motion of the RLG cavity due to local geophysical and instrumental effects, where rotations and tilts are mixed together,  thus making possible to reconstruct $\omega_L$ in Eq.~\ref{basicTheory}.  
 Following  Eq.~\ref{basicTheory}, effects of changes in the projection are evaluated by the product between $\delta\theta_L$ and the total rotation velocity $(\Delta\omega_3 + \omega_L+\overline{\Omega}_\oplus + \Omega_{GR})$. The latter must be determined by iterating the linear regression procedure, starting with an average estimation of the angular velocities and leading to convergence typically in a couple of iterations. The projection through $\cos\theta$ of the term $PM$, accounting for the combined effects of polar motion and Annual and Chandler wobbles, can also be determined following a similar iterative procedure. 
 Data from a temperature sensor are included in the linear regression, although their effect is rather small.
It has been checked that pressure variations, measured by dedicated probes, do not play a significant role.
The Sagnac angular frequency ($\omega_s = 2\pi f_s$) can be determined from the measured beat note and the output power of the laser beams counter-propagating in the RLG cavity, according to the recently developed approach described in \cite{DiVirgilio2019,DiVirgilio2020}: the first step of the analysis  removes the backscatter noise providing an initial evaluation of the Sagnac angular frequency $\omega_{s0}$, the second step eliminates laser dynamics. Briefly, the approach accounts for the occurrence of technical issues such as, those related to dark currents in the photodetectors, those associated with the laser operation in the two counter-propagating beams, and the related losses, via different correcting terms, denoted as $\omega_{\xi}$, $\omega_{ns1}$, $\omega_{ns2}$, and specifically defined form factors. 
Since the linear regression procedure leading to $F_{geo}$ involves both AC and DC terms, the latter including the above mentioned $\omega_\xi$, $\omega_{ns1}$, $\omega_{ns2}$, and $\overline{\Omega}_\oplus+\Omega_{GR}$, a cross calibration of $f_s$ with IERS data is needed, which is accomplished at a single, arbitrarily chosen, data point. It has been checked that results do not depend on the choice of the calibration point.
In the calculation, a scale factor $S = 5.6890\times 10^6$ rad/s/Hz is used, whose accuracy is dictated by the machining precision (tenths of millimeter) of the monument supporting the RLG. $S$ and the projection angle are multiplicative factors, which cannot be distinguished each other. It is assumed that the uncertainty in the proportionality constant is mainly due to the orientation of the cavity, in particular to the deviations of its plane with respect to the horizontal one. The latitude of the underground laboratory is $0.7409$ rad, whereas the analysis indicates a value of $0.7420$ rad, corresponding to an extra $\simeq 1$ mrad in the absolute orientation of the ring cavity.
\\
The general scheme of the analysis is shown in Fig.~\ref{Scheme}, which follows the approach that we have recently developed.\cite{DiVirgilio2019, DiVirgilio2020}
 Two distinct methods have been applied to the linear regression procedure. In the first one, all available data are used at once. In the second method, linear regression is carried out on moving windows consisting of three days, with an half day overlap before and after each window. 
 Parameters resulting from the linear regression of the whole data set are used as the initial guess in the moving window method, in particular the guess of $\omega_s$, accounting for the contribution of the terms describing the projection by the angle $\delta\theta_L$ in Eq.~\ref{basicTheory}. In each window the cross-calibration procedure is repeated.
 Bandwidth of the analysis corresponds to a temporal resolution of 600 s.
Two independent and very different sets of data have been analysed: 30 days from June 16, 2018, and 70 days from October the first, 2019. Data pertaining to  2018 showed higher contrast in the Sagnac interferogram and higher duty cycle, since the temperature was a factor ten more stable than for the 2019 data set. Approximately $85\%$ and $79\%$ of data have been selected, respectively.
\\
Figure \ref{fig:sub70} demonstrates the effect of correcting data of the 2019 set for  laser systematics (a) and for $\delta \theta_L$, $\omega_L$ (b); effects of temperature variations are not shown since they are at the level of $10~\mu$Hz, and not well visible in the graphs. Panel (c)  shows the final result, $F_{geo}$, as obtained with the single-fit method, compared with $F_{IERS}$. The Annual and Chandler wobbles effect, reflected in the slow trend of data, is well reproduced in the single-fit. On the contrary, the approach turns out not sufficient for retrieving the daily polar motion, which is barely visible in panel (c). 
The  2018 data set shows a similar behaviour: the laser systematics and the tilt-meter corrections $\delta\theta_L$ and $\omega_L$ take into account most of the disturbances. 
Remarkably, despite the qualitative difference between the two data sets, the evaluations exhibit very similar uncertainties: Root Mean Square (RMS) errors of $10.8~\mu$Hz and $10.7~\mu$Hz are obtained. Typically, the covariance test ANOVA gives  high F-statistics versus constant model and  p-value close to zero.
The inadequate reproduction of the daily polar motion achieved with the single-fit method is not surprising, since polar motion is a periodic signal with a null mean value for long time average. Moreover, the model adopted in the analysis can turn out oversimplified, since it accounts only for  daily variations, while it is well known that other very similar signals are present in that frequency bandwidth.\cite{Tercjak2017} The analysis is clearly improved when the fit is carried out over the moving windows. In particular, the daily polar motion is correctly recovered, as demonstrated in Fig.~\ref{fig:comp70}, referred to 2019 (a) and 2018 (b) data sets.
 It must be noted that the observed daily variation has been carefully checked not to stem from disturbances, e.g., daily noise due to anthropic activity, which is strongly suppressed in the underground laboratory. Moreover, although data statistics is not large enough to precisely discriminate the frequency of the daily polar motion from tides, it has been verified that the reconstructed signal is constantly in phase with the one expected for the polar motion.
Remarkably, the approach shows also predictive capabilities. Indeed, by fitting on the moving window over a certain sub-set of available data, the behaviour for a subsequent sub-set can be reliably predicted. Reconstructed signal in Fig.~\ref{fig:comp70}(a) for the last two days, marked in red, has been produced by using the linear regression in the previous period. 
\\
 $\Delta\omega_3$, i.e., the low frequency variations of the Earth rotation rate connected
to changes of LOD and to zonal tides, is provided by IERS with two options, with or without zonal tides. Its estimate based on GINGERINO data can be obtained by comparing results obtained by including, or not,  its contribution in the linear regression procedure. 
Figure \ref{fig:lod1} shows the evaluated $\Delta\omega_3$ for the  2019 data set using both the mentioned options.
Comparison of the results obtained from GINGERINO and IERS data leads to a standard deviation of the residual $\simeq 0.2$ prad/s. The agreement is even better when $\omega_{ZT}$ is evaluated by arranging results of different analyses. 
We remark that, in the case $F_{geo}$ is affected by unidentified disturbances not included in the model, their effect would be subtracted in the evaluation of $\Delta \omega_3$, being them present in both evaluations.
 Found values of $\Delta\omega_3$ are close to the noise floor of the apparatus. In this case the method is not predictive and provides mainly compatibility between model and available data. Moreover, to check the validity of the analysis, synthetic signals, either chirped or periodic at low frequency, of the same order of magnitude of $\Delta\omega_3$ have been used. When added to both the data and the model, such synthetic signals have been correctly identified.  In general the analysis leads to larger errors when signals are inserted in the model only, but not always a signal present in the model and not added to the data is recognized as a fake. To improve the validity of the analysis for so small signals it will be necessary to improve both calibration of the RLG and the model, in order to achieve a more effective rejection of local disturbances.
 \\
The feasibility of the GR test has been checked comparing the results obtained with different models: the difference of the average values of $F_{IERS}$ including and neglecting $\Omega_{GR}$ is consistent with the expected GR shift.
Since $\Omega_\oplus$ is rather stable in time, by analogy with clock frequency analysis the quantity relevant for assessing the quality of the approach and of the produced results is the Allan variance, in particular the Modified Allan Deviation (MAD),\cite{Allan} which provides also indications on the nature of the noise limiting the measurement accuracy.
Figure \ref{fig:MADl} shows the MAD calculated for $\omega_{s0}$, for $F_{geo}$ and for the residuals, defined as $(F_{geo} - F_{IERS})$. The different MAD found for $\omega_{s0}$ in the two data sets confirms the  already mentioned differences between them. The long time behaviour very well follows the IERS data for both data sets, as highlighted by the consistent decrease  of the residual MAD with time. The peak around $3-4 \times 10^4$ s  indicates the occurrence of other signals at that frequency that have not been included in the model or cannot be retrieved by using only tilt-meter data. 
The behaviour shown in Fig.~\ref{fig:MADl} suggests that shot-noise plays a dominant role in the measurement, while being larger than the theoretical shot-noise, evaluated for GINGERINO below 1 prad/s in 1 s measurement time. 
Tilt-meter is not the best suited instrument to monitor $\delta\theta_L$ since it provides it  with respect to the local vertical, while the analysis would need changes with respect to the rotation axis of $\vec{\Omega}_T$. We remark that the availability of an array of RLGs, as foreseen in the GINGER project, would make not necessary using the tilt-meter data. In particular, the problem of evaluating $\delta\theta_L$ would be fully solved by using a pair of RLGs, with one oriented at the maximum Sagnac signal, or independently measuring the relative angle between the two RLGs.\cite{Angela2017}
\\
It is therefore demonstrated that with a minimum square linear regression procedure, taking into account the expected $F_{IERS}$, the laser dynamics, and the environmental monitors, it is possible to recover geodetic effects from GINGERINO data. In particular,  $F_{geo}$ reproduces all its main features such as, Annual and Chandler wobbles, daily polar motion, 
and the very low frequency contribution ($\Delta\omega_3$) due to LOD and zonal tides. Uncertainties are below 1 $\mu$Hz (equivalent to an error in the evaluated angular rotation rate of $1.7\times10^{-13}$ rad/s) in a bandwidth corresponding to 600 s measurement time. 
The residuals  show the occurrence of other daily and sub-daily signals, which are absent in the used model containing only main contributions to the observed quantities. Remarkably, their MAD exhibits a decrease with the measurement time similar to the one shown by shot-noise, and eventually drops down to 18 prad/s Hz$^{-1/2}$, approximately ten times above the theoretical shot-noise of GINGERINO, which could be probably further reduced by getting rid of tilt-meter signals in the linear regression, as enabled by an array of RLGs. \\
The analysis demonstrates the very high sensitivity which can be obtained by using RLGs, and point to the necessity of identifying and subtracting systematics associated with non linear laser dynamics.
The underground location provides several advantages and sensitivity is shown adequate for the purpose of the GINGER project,\cite{Angela2017} where an array of RLGs, with improved heterolithic structure and independent calibration strategies, are foreseen. 
\\
Cross calibration of presently available GINGERINO data with IERS
is effectively accomplished 
 for $\Omega_\oplus$, paving the way for a reliable use of RLG data in the investigation of GR effects with Earth-based measurements. 
\section*{Acknowledgments}
We thank the Gran Sasso staff in support of the experiments, particularly Stefano Gazzana. We thank Ulli Schreiber for his continuous support in discussion and in favour to RLG. We thank Giancarlo Cella and Roberto Devoti for useful discussions.

\end{document}